# Structure and Depletion at Fluoro- and Hydro-carbon/Water Liquid/Liquid Interfaces

Kaoru Kashimoto[1,2], Jaesung Yoon[1], Binyang Hou[1], Chiu-hao Chen [1], Binhua Lin[3], Makoto Aratono[2], Takanori Takiue[2], and Mark L. Schlossman[1,*]

[1]*Department of Physics, University of Illinois at Chicago, Chicago, IL 60607.*

[2]*Department of Chemistry and Physics of Condensed Matter, Graduate School of Sciences, Kyushu University, Fukuoka 812-8581, Japan.*

[3]*The Center for Advanced Radiation Sources, University of Chicago, Chicago, IL 60637.*

Abstract:

The results of x-ray reflectivity studies of two oil/water (liquid/liquid) interfaces are inconsistent with recent predictions of the presence of a vapor-like depletion region at hydrophobic/aqueous interfaces. One of the oils, perfluorohexane, is a fluorocarbon whose super-hydrophobic interface with water provides a stringent test for the presence of a depletion layer. The other oil, heptane, is a hydrocarbon and, therefore, is more relevant to the study of biomolecular hydrophobicity. These results are consistent with the sub-angstrom proximity of water to soft hydrophobic materials.

The formation of a vapor-like depletion layer at the interface between an aqueous solution and a hydrophobic material would have important consequences for many biological, chemical, and environmental processes. Such a layer would limit water molecules, and solutes in the water, from being positioned immediately adjacent to a hydrophobic material. This would affect dynamic processes such as protein folding and self-assembly in which hydrophobic regions are briefly exposed to water, as well as the wetting of aqueous solutions on many environmental and industrially important surfaces.

Early theoretical studies indicated that water maintains a hydrogen bond network surrounding small hydrophobic solutes, though this network would be stretched with increasing solute size [1]. For an infinite radius solute, or equivalently for a planar interface, Stillinger predicted the existence of a depletion layer that is essentially a water-vapor interface that forms near the hydrophobic plane [1]. Recent theory and computer simulations suggest the presence of a depletion layer whose thickness is on the order of a few angstroms for solutes of several nanometers or larger radius [2-6]. Attractive interactions between the hydrophobic material and water are expected to thin the depletion layer [7]. Recent simulations of planar interfaces that include attractive interactions have suggested that a master curve describes the variation of the depletion layer thickness with surface hydrophobicity [6].

Numerous experimental studies have provided conflicting evidence for and against the presence of a depletion layer [8-19], though recent x-ray and neutron scattering studies from the interface between water and a hydrophobic coating on a solid provide evidence for a depletion layer with thickness of a few angstroms [13-15, 17]. Since biological hydrophobic/aqueous interfaces are often soft, there is a need for studies of soft and well-defined hydrophobic/aqueous interfaces. Recent optical ellipsometry studies of liquid/liquid interfaces were analyzed by one of two models that yielded a range of depletion layer thickness values that varied from 0.3 Å to 3 Å for the same sets of data [12]. Also of interest are vibrational sum frequency spectroscopy studies that demonstrate much weaker hydrogen bonding in the region of the water/hydrophobic liquid interface than at the water/vapor interface [20]. These data suggest that a bulk-like water/vapor interface does not form near a water/hydrophobic liquid interface. However, the lack of depth sensitivity in the spectroscopy measurements preclude them from providing a decisive argument for or against the existence of depletion layers. Here, we present x-ray reflectivity studies of two oil/water (liquid/liquid) interfaces. One of these interfaces, the perfluorohexane/water interface, is super-hydrophobic and, therefore, provides a stringent test of the depletion layer prediction. The other is a hydrocarbon (heptane)/water interface that is more relevant to the study of biomolecular hydrophobicity. The materials used for these studies were extensively purified and their purity was tested with tensiometry and gas-liquid chromatography (see supplementary information [21] for a detailed description).

Teflon and other fluorocarbon materials are strongly hydrophobic. The super-hydrophobicity of the interface between liquid perfluoro-$n$-hexane ($CF_3(CF_2)_4CF_3$) and water at 23.5 °C is evident in Fig. 1 which shows a spherical drop of water in equilibrium at the perfluorohexane/air interface. The dihedral angle is 180° to within our measurement accuracy ($\pm 1°$). This is consistent with the large, negative spreading coefficient of −110.7 mN m$^{-1}$ determined by our measurements of interfacial tension, which indicates that water does not wet perfluorohexane (see supplementary information [21]).

X-ray reflectivity probes the electron density variation with interfacial depth. Figure 2 illustrates reflectivity data $R(Q_z)$ as a function of the wave vector transfer $Q_z$ (normal to the interface) from a flat, circular perfluorohexane/water interface of 70 mm diameter [21]. The reflectivity was determined by measurements of the reflected x-ray intensity normalized by the incident intensity, after subtraction of background scattering [21-23]. The reduction of the measured reflectivity below the calculated Fresnel reflectivity $R_F(Q_z)$ expected from an ideal, smooth interface (see Fig. 2) can be described by [24]

$$R(Q_z) \approx R_F(Q_z) \exp(-Q_z Q_z^T \sigma^2), \qquad (1)$$

where $Q_z^T \approx \sqrt{Q_z^2 - Q_c^2}$ is the wave vector transfer in the lower (*e.g.*, perfluorohexane) phase and the critical wave vector transfer for total x-ray reflection is calculated to be $Q_c = 0.01458$ Å$^{-1}$ for perfluorohexane/water and $Q_c = 0.01169$ Å$^{-1}$ for heptane/water interfaces [21]. The interfacial width $\sigma$ describes the crossover from one bulk phase to the other via an error function, $erf(z/\sqrt{2}\sigma)$ (see supplementary information [21]). A two parameter fit of the reflectivity data yields $\sigma = 3.4\pm0.2$ Å and $Q_{\text{offset}} = 0.0004$ Å$^{-1}$, where the latter is an additive offset in $Q$ that represents a typical misalignment of the x-ray instrument. This fit produces the solid line that is in good agreement with the data in Fig. 2. Although Eq.(1) does not include the resolution dependent contribution from capillary waves, this is a small effect for these experiments (because $\eta \le 0.2$, see [25, 26]).

Capillary waves at the interface, driven by thermal energy, will scatter x-rays out of the specularly reflected beam and reduce the measured reflectivity below the Fresnel reflectivity [27]. Hybrid capillary wave theory describes the total interfacial width as $\sigma^2 = \sigma_{\text{cap}}^2 + \sigma_{\text{int}}^2$ [28]. The intrinsic width $\sigma_{\text{int}}$ represents the effect of interfacial molecular ordering and $\sigma$ is a result of the intrinsic profile fluctuating according to the capillary wave spectrum. Capillary wave theory predicts $\sigma_{\text{cap}} = 3.37$ Å (see supplementary information [21]), in agreement with our measurement of $\sigma$. Therefore, $\sigma_{\text{int}}$ is small, suggesting that the intrinsic molecular ordering at the interface is weak.

The high accuracy of the data in Fig. 2 allows us to present them in the inset to Fig. 2 as reflectivity normalized to the Fresnel reflectivity $R(Q_z)/R_{\text{F}}(Q_z)$. This reveals small but significant deviations of the data from the fit to Eq.(1). As described shortly, these deviations cannot be explained by a depletion layer (or by any single layer, see supplementary information [21]). Although these data cannot uniquely specify the molecular origin of this effect, they can be modeled by multiple layering of perfluorohexane molecules at the interface. The fitting shown in the Fig. 2 inset used an intrinsic profile $\langle\rho(z)\rangle_{\text{int}} = \rho_{\text{f}} + A\left[C - \cos(2\pi z/l_{\text{osc}} + \phi)\right]\exp(-z/l_{\text{dec}})$, where one period of the cosine represents a layer of perfluorohexane molecules. The fraction of perfluorohexane molecules that form a layer decreases with increasing distance from the interface according to the exponential decay length $l_{\text{dec}}$ (supplementary information [21]). This modeling suggests that two to three layers of perfluorohexane exist at the interface, though the maximum density of these layers is only 3% above the bulk liquid perfluorohexane density. Given the rigid, nearly cylindrical shape of perfluorohexane molecules, this layering is reminiscent of smectic multi-layering that occurs at the surface of liquid crystals [29].

Figure 3 demonstrates that the reflectivity data from the perfluorohexane/water interface is inconsistent with the presence of a depletion layer. The interface is modeled by back to back water/vapor and vapor/oil interfaces that are separated by a distance $D_{dep}$ (Fig. 3 inset). The electron density profile $\langle \rho(z) \rangle_{xy}$ of this interface is given by

$$\langle \rho(z) \rangle_{xy} = \frac{1}{2} \rho_{water} \left[ 1 - \mathrm{erf}\left( \frac{z}{\sqrt{2}\sigma} \right) \right] + \frac{1}{2} \rho_{oil} \left[ 1 + \mathrm{erf}\left( \frac{z - D_{dep}}{\sqrt{2}\sigma} \right) \right]. \tag{2}$$

This model represents a depletion layer of vapor-like density and thickness $D_{dep}$ sandwiched between bulk water and oil. The water/vapor and vapor/oil interfaces fluctuate with capillary waves, as represented by the error function in Eq.(2). We let $\sigma = 3.4$ Å, as determined by the x-ray measurements. It could be argued that a vapor-like depletion layer will have interfacial widths characteristic of the bulk water/vapor and vapor/oil interfaces, but the data cannot be fit with such values for the widths [21]. The data in Fig. 3 are sensitive to the presence of depletion layers of 0.2 Å thickness and exclude any such layer of this thickness or greater. This is generally much thinner than depletion layers previously reported.

A similar reflectivity measurement and analysis has been carried out for the interface between water and a hydrocarbon, hydrophobic liquid – heptane $(CH_3(CH_2)_5CH_3)$. The normalized reflectivity $R(Q_z)/R_F(Q_z)$ measured from this interface is shown in Fig. 4. A one parameter fit to Eq.(1) is excellent, with a value of the interfacial width $\sigma = 4.2 \pm 0.2$ Å. The small deviations observed in the fit to Eq.(1) for the perfluorohexane/water interface (Fig. 2 inset red line) are absent. The heptane molecules are flexible and it is unlikely that they would form smectic layers as our data suggest for the rod-like perfluorohexane molecules. This is supported by our interfacial tension measurements as a function of temperature of a similar system, the hexane/water interface, that indicate that it is slightly more disordered than the perfluorohexane/water interface (the excess interfacial entropy, $\Delta s_{perfluorohexane/water} = 0.070$ mJ $K^{-1}$ $m^{-2}$ and $\Delta s_{hexane/water} = 0.083$ mJ $K^{-1}$ $m^{-2}$ at 25 ºC, see supplementary information [21]).

The interfacial width predicted from capillary wave theory for the heptane/water interface is 3.44 Å, which is different from the measured value of 4.2±0.2 Å. Figure 4 demonstrates that this difference is not due to a depletion layer because the presence of a depletion layer would increase the reflectivity above the measured values, as well as above the values predicted by capillary wave theory. Equation (1) indicates that the larger value of the measured width has the opposite effect. It acts to decrease the reflectivity below the values predicted by capillary wave theory.

Measured values of interfacial width larger those of the capillary wave prediction have been reported for the liquid alkane/water interface for a range of alkane chain lengths from 6 to 22 carbons [30]. For the reasons just described, the presence of a depletion layer (as in Eq.(2)) cannot explain these variations. A larger interfacial width could be due to the presence of molecular ordering at the interface that leads to an intrinsic width $\sigma_{int}$ such that the total width $\sigma^2 = \sigma_{cap}^2 + \sigma_{int}^2$ is increased. A phenomenological explanation suggested that the intrinsic width was given by the shorter of two relevant length scales that describe alkane ordering: the radius of gyration of the alkane and the bulk correlation length in the alkane liquid [30]. Although this explanation described quantitatively the linear increase in $\sigma$ with chain length for alkanes of length 6 to 16 carbons, and the constant value for longer alkanes, it did not fully justify that these length scales should characterize the increase in interfacial electron density required to explain the data.

Recent simulations of the hydrophobic/water interface utilized a planar solid surface with a variable hydrophobicity to predict values of the depletion layer thickness to be a few angstroms [6]. Extrapolation of the master curve that resulted from these simulations to a super-hydrophobic interface, such as our perfluorohexane/water interface , would predict a depletion layer thickness of ~3Å, in contrast to the upper bound of 0.2Å determined by our data. With regard to our results on the heptane/water interface, calculations have suggested that attractive van der Waals interactions between a hydrocarbon oil and water will thin the depletion layer to nearly zero thickness [7]. Chandler has suggested that the weakness of the attractive van der Waals forces, on the order of $k_BT$, will result in enhanced interfacial fluctuations in the thickness of the depletion layer [31]. Such fluctuations are not included in the capillary wave theory. If present, we anticipate that the interfacial width measured by x-ray reflectivity would increase because x-rays would be scattered by the fluctuations. Development of a theory of depletion layer fluctuations, that describes the height-height correlation function of the interface, may help resolve the issue of enhanced interfacial width at the alkane/water interface.

K.K. acknowledges a fellowship from the Japan Society for the Promotion of Science. M.L.S. acknowledges support from NSF-CHE . ChemMatCARS is supported by NSF-CHE, NSF-DMR, and the US Department of Energy (DOE-BES). The Advanced Photon Source at Argonne National Laboratory and the National Synchrotron Light Source at Brookhaven National Laboratory are supported by the US DOE-BES.

**Figure legends**

Figure 1  Water drop placed at the perfluorohexane/vapor interface has a dihedral angle of ~180°, indicating that perfluorohexane is super-hydrophobic.

Figure 2  X-ray reflectivity from the perfluorohexane/water interface at 23.5 ℃ as a function of wave vector transfer normal to the interface. Solid line is the fit to the capillary wave theory. Dashed line is the Fresnel reflectivity, $R_F$. Error bars on the data are smaller than the symbols. The point at $Q_z = 0$ is the direct (not reflected) beam that is used to normalize the reflectivity.   The inset is x-ray reflectivity normalized to the Fresnel reflectivity.  The small variation from the capillary wave theory fit (red dashed line) indicates the presence of additional weak structure.  The good fit to the black line suggests multi-layering of perfluorohexane at the interface.

Figure 3  X-ray reflectivity normalized to the Fresnel reflectivity analyzed by the depletion layer model for the perfluorohexane/water interface.  Symbols are experimental results.  Solid line is a fit to Eq.(1).  Dashed lines are determined by Eq. (2) with depletion distances $D_{dep} = 0.5$ Å to 3 Å in steps of 0.5 Å (bottom to top) with an interfacial width of $\sigma = 3.4$ Å. The inset shows the electron density profile normalized to the bulk water, $\rho(z)/\rho_w$, which is used to calculate the corresponding X-ray reflectivity. The solid lines show the water and perfluorohexane densities separately for the case $D_{dep} = 3$ Å, where there is a 3 Å separation between the midpoints of these lines (indicated by the vertical lines). The dashed lines are for $D_{dep} = 0$ Å to 3 Å in steps of 0.5 Å (top to bottom), where $D_{dep} = 0$ Å corresponds to the solid line in the main figure (which is the same as the red dashed line in the inset to Fig. 2).

Figure 4  X-ray reflectivity normalized to the Fresnel reflectivity analyzed by the depletion layer model for the heptane/water interface.  Symbols are experimental results.  The point at $Q_z = 0$ is the direct (not reflected) beam that is used to normalize the reflectivity.  Solid line is a fit to Eq.(1). Long dashed line is the capillary wave fit with $\sigma_{cap} = 3.44$ Å.  Short dashed lines are determined by Eq. (2) with depletion distances $D_{dep} = 0.5$ Å to 3 Å in steps of 0.5 Å (bottom to top) with an interfacial width of $\sigma_{cap} = 3.44$ Å.

**Figures**

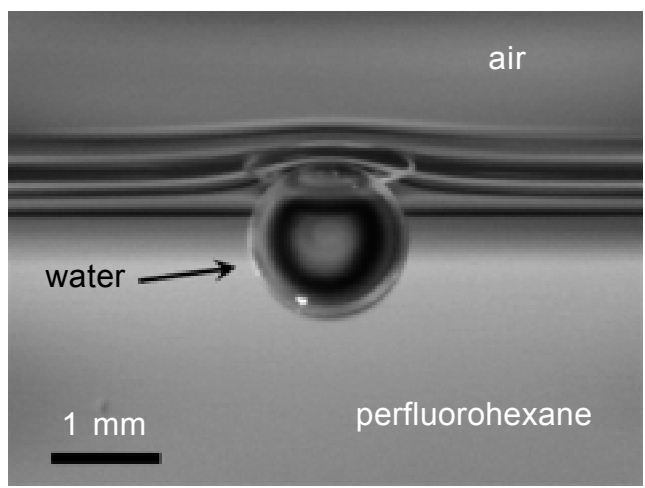

Figure 1

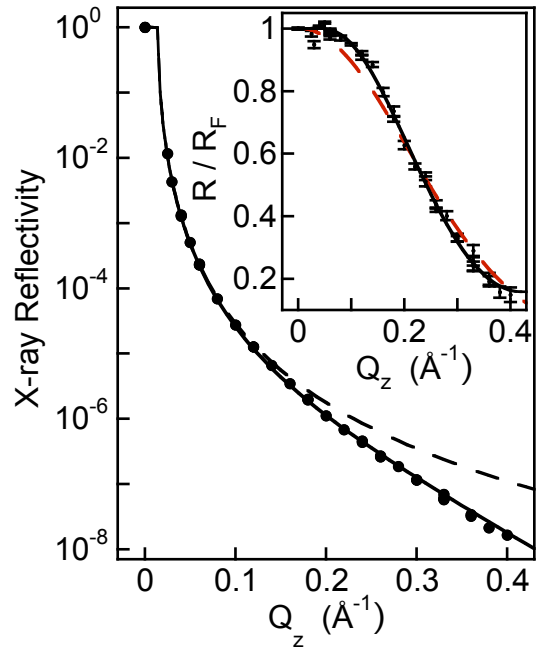

Figure 2

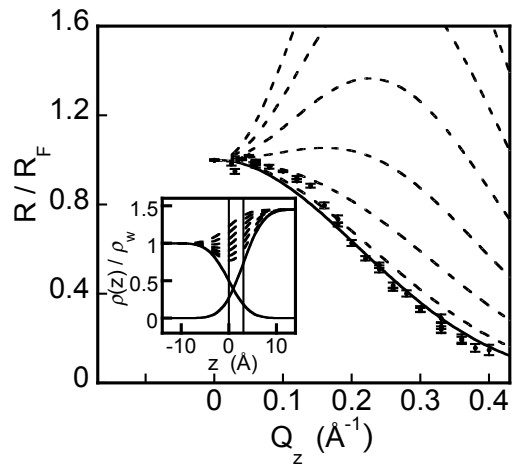

Figure 3

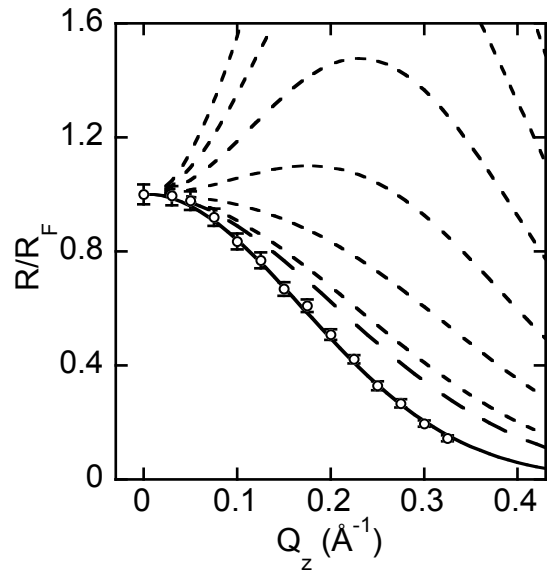

Figure 4

Supplementary Information for

**Structure and Depletion at Fluoro- and Hydro-carbon/Water**

**Liquid/Liquid Interfaces**

Kaoru Kashimoto[1,2], Jaesung Yoon[1], Binyang Hou[1], Chiu-hao Chen [1], Binhua Lin[3], Makoto

Aratono[2], Takanori Takiue[2], and Mark L. Schlossman[1,*]

*[1]Department of Physics, University of Illinois at Chicago, Chicago, IL 60607.*

*[2]Department of Chemistry and Physics of Condensed Matter, Graduate School of Sciences,*

*Kyushu University, Fukuoka 812-8581, Japan.*

*[3]The Center for Advanced Radiation Sources, University of Chicago, Chicago, IL 60637.*

## Materials

Tetradecafluorohexane, (perfluoro-*n*-hexane; FC6) was purchased from Aldrich Chemical Co. Inc. (99%) and distilled at 57 ℃ under atmospheric pressure after extracting hydrophilic impurities by triple extraction with water (1:1 volume ratio). The purity was checked by observing no time dependence of the water/FC6 interfacial tension and by gas-liquid chromatography. Heptane was purchased from Fluka and purified by passing it several times through a column of basic alumina. Its purity was also checked by the time-dependence of the tension of a newly formed heptane/water interface. Triply distilled water was used for interfacial tension measurement and wetting observations; the second and third stages were carried out with dilute alkaline permanganate solution. The samples studied by x-ray reflectivity measurements used water produced by a Millipore Milli-Q system. The density of FC6 was measured as a function of temperature under atmospheric pressure by using an Anton Paar DMA 60/602 vibrating tube densimeter [1, 2]. The density of FC6 is 1.6748 g cm$^{-3}$ at 23.5 ℃ (296.65 K) and 1.6704 g cm$^{-3}$ at 25 ℃ (298.15 K), in agreement with literature values [3]. Density values for water were taken from the literature [4].

## Interfacial Tension

The interfacial tension at the air/water and air/FC6 interfaces was measured at 25 ℃, and the tension at the water/FC6 interface was measured as a function of temperature, under atmospheric pressure by the pendant drop method [5]. The measurement at the air/water interface was first carried out in the absence of FC6, then the tension was measured after the water reached equilibrium with a drop of FC6 placed in the bottom of the water container. The estimated experimental error of interfacial tension was ± 0.05 mN m$^{-1}$. The tension of the heptane/water interface was measured using the Wilhelmy plate method.

The interfacial tensions measured at water/FC6 interface $\gamma^{W/O}$ are plotted against temperature $T$ together with the plots of pure hexane (C6)/water interface in Supplementary Figure 1. The interfacial tensions of both interfaces decrease monotonically with increasing $T$, but their slopes are slightly different. The temperature dependence of interfacial tension gives the entropy of interface formation $\Delta s$ by $\Delta s = -(\partial \gamma / \partial T)_P$. The $\Delta s$ value is smaller for water/FC6 interface (0.070 mJ K$^{-1}$ m$^{-2}$ at 25 °C) than for C6/water interface (0.083 mJ K$^{-1}$ m$^{-2}$ at 25 °C).

Supplementary Figure 1. Interfacial tension versus temperature curves: (open circles)

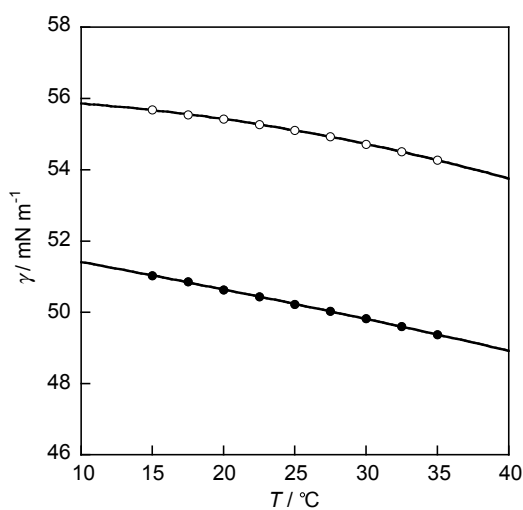

water/perfluorohexane interface; (dots) hexane/water interface. The lines are guides to the eye.

## Wetting and Spreading Coefficient

Wetting behavior is often utilized to assess the hydrophobicity of solid substrate. Following this treatment, a snapshot of a water drop on FC6 surface was taken with a CCD camera and is shown in Figure 1 of the main paper. The dihedral angle $\theta$ for the water lens is visually very close to 180°, which enables us to conclude that the hydrophobicity of FC6 is very high. It is noted that the meniscus appears to be spread over the floating water lens. This behavior is governed by a balance of interfacial tensions of air/water ($\gamma^{A/W}$), air/FC6 ($\gamma^{A/O}$) and water/FC6 ($\gamma^{W/O}$) interfaces [6, 7]. Their values are $\gamma^{A/W} = 66.87$, $\gamma^{A/O} = 11.26$, and $\gamma^{W/O} = 55.11$ mN m$^{-1}$ at 25 °C. These values are comparable with the case of Teflon; $\gamma^{A/Teflon} \approx 19$ and $\gamma^{W/Teflon} \approx 50$ mN m$^{-1}$ [8]. It should be noted that the measured interfacial tension at the air/water interface in the absence of FC6, $\gamma^{A/W\text{-}o}$ is 71.96 mN m$^{-1}$. Since the solubility of FC6 in water is negligibly small, the difference between $\gamma^{A/W}$ and $\gamma^{A/W\text{-}o}$ seems to be caused by the adsorption of FC6 at the air/water interface. The wetting observation is substantiated by calculating the spreading coefficient $S$ defined by $S = \gamma^{A/O} - (\gamma^{A/W} + \gamma^{W/O})$. The value of $S$ is $-110.72$ mN m$^{-1}$, which indicates non-wetting of water on FC6 surface. However, the condition $\gamma^{A/W} > (\gamma^{A/O} + \gamma^{W/O})$ holds and therefore FC6 wets the water surface.

**X-ray Sample Cell**

The x-ray sample cell used for measurements from the C7/water interface was described in Ref. 11. For X-ray reflectivity measurements of the FC6/water interface, the two liquids were contained in a vapor tight round glass sample cell of 70 mm diameter (Supplementary Figure 2). To reduce the meniscus due to the contact of liquids with glass wall, a polytetrafluoroethylene (Teflon) strip was pressed to the glass surface with a spring made of stainless steel shim stock. The liquid/liquid interface was pinned by the top edge of the Teflon strip. Fine-tuning of the interfacial flatness was accomplished by controlling the volume of FC6 (the lower phase) with a syringe connected to the side tube of sample cell. The interfacial area, where both water and FC6 phases meet, was 38.5 cm$^2$ with X-rays penetrating through the upper water phase. Samples were equilibrated at $23.5 \pm 0.5$ °C prior to the measurements.

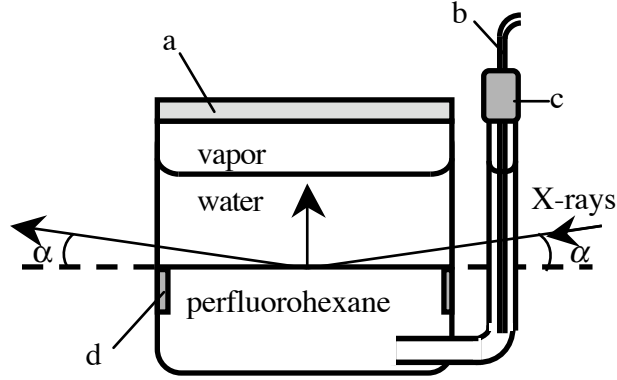

Supplementary Figure 2. Schematic illustration of glass sample cell: (a) glass cover; (b) tube connected to syringe; (c) rubber cap; (d) Teflon strip to capture meniscus of interface. The kinematics of interfacial X-ray reflectivity is also indicated: $\mathbf{k_{in}}$ is the incoming X-ray wave vector, $\mathbf{k_{scat}}$ is the scattered wave vector, $\alpha$ is the angle of incidence and reflection, and $\mathbf{Q}_z$ is the wave vector transfer normal to the interface.

**Capillary Wave Theory for the Interfacial Width**

The BLS capillary wave theory[9] model for the interface corresponds to an error function profile for the electron density averaged over the plane of the interface, $\langle \rho(z) \rangle$, given by

$$\langle \rho(z) \rangle = \frac{1}{2} \rho_{\mathrm{w}} \left[ 1 - \mathrm{erf}\left( \frac{z}{\sqrt{2}\sigma} \right) \right] + \frac{1}{2} \rho_{\mathrm{O}} \left[ 1 + \mathrm{erf}\left( \frac{z}{\sqrt{2}\sigma} \right) \right], \quad \text{with} \quad \mathrm{erf}(z) = \frac{1}{\sqrt{\pi}} \int_0^z e^{-t^2} \mathrm{d}t \;, \qquad \text{(S1)}$$

where $\rho_{\mathrm{w}}$ and $\rho_{\mathrm{O}}$ represent the electron densities of bulk water and bulk oil, and $\sigma$ is the interfacial width. The interfacial width $\sigma$ is considered to be the combination of two different contributions, the intrinsic profile $\sigma_{\mathrm{int}}$ and the capillary wave contribution $\sigma_{\mathrm{cap}}$, by $\sigma^2 = \sigma_{\mathrm{cap}}^2 + \sigma_{\mathrm{int}}^2$ [10]. The BLS capillary wave theory [9] predicts $\sigma_{\mathrm{cap}}$ to be

$$\sigma_{\mathrm{cap}}^2 = \frac{k_{\mathrm{B}}T}{2\pi} \int_{q_{\mathrm{min}}}^{q_{\mathrm{max}}} \frac{q}{\gamma^{\mathrm{W/O}} q^2 + \Delta\rho_{\mathrm{m}} g} \mathrm{d}q \approx \frac{k_{\mathrm{B}}T}{2\pi \gamma^{\mathrm{W/O}}} \ln\left( \frac{q_{\mathrm{max}}}{q_{\mathrm{min}}} \right), \qquad \text{(S2)}$$

where $k_{\mathrm{B}}T$ is the Boltzmann constant times the temperature, $\gamma^{\mathrm{w/o}}$ is either the perfluorohexane/water interfacial tension at 23.5 °C (55.20 mN m$^{-1}$) or the the heptane/water interfacial tension at 25.0 °C ($\gamma^{w/o} = 51.7$ mN m$^{-1}$), $\Delta\rho_m$ is the mass density difference of the two bulk phases, $g$ is the gravitational acceleration, $q_{min} = (2\pi/\lambda)\,\Delta\beta\,\sin\alpha$ and $\Delta\rho_m g \ll q_{min}^2$. Here the angular acceptance of the detector $\Delta\beta = 3.5$ x $10^{-4}$ rad for the perfluorohexane/water measurement and 8.9 x $10^{-4}$ rad for the heptane/water measurement, and $\alpha$ is the largest reflection angle. The variable $q$ is the in-plane wave vector of the capillary waves. The limit, $q_{max}$, is determined by the cutoff for the smallest wavelength capillary waves that the interface can support. We have chosen $q_{max} = 2\pi/D$ Å$^{-1}$, where D is an approximate molecular size (5 Å for heptane and 6 Å for perfluorohexane). Note that $\sigma_{cap}$ depends on $q_{max}$ logarithmically and is not very sensitive to its value.

**X-ray Measurements**

X-ray reflectivity of the water/FC6 interface was measured at the ChemMatCARS beamline 15-ID at the Advanced Photon Source (Argonne National Laboratory, USA) and the reflectivity of the water/heptane interface was measured at Beamline X19C at the National Synchrotron Light Source (Brookhaven National Laboratory). The measurements were carried out with a liquid surface instrument and measurement techniques described in detail elsewhere [11, 12]. The kinematics of reflectivity is also illustrated in Supplementary Figure 2. For specular reflection, the wave vector transfer, $Q = k_{scat} - k_{in}$, is only in the $z$-direction, normal to the interface; $Q_z = (4\pi/\lambda)\sin\alpha$, where $\lambda = 0.41188 \pm 0.00005$ Å is the X-ray wavelength for water/FC6 ($\lambda = 0.825 \pm 0.002$ Å for water/heptane) , and $\alpha$ is the incident angle. To set the incident beam size, an incident slit (typically 15 μm x 3 mm, vertical x horizontal) was placed 68 cm upstream of the liquid sample. For the measurement at X19C, two slits were used which produced a vertical divergence of 20 microrads. An ion chamber (or kapton film that scattered x-rays into a scintillator detector for the water/heptane measurement) was placed before the sample to measure the incident X-ray flux. The sample was followed by a slit with a vertical gap of typically 0.3 mm (4 mm horizontal gap) to reduce the background scattering and a scintillator detector was preceded immediately by a slit with a vertical gap of typically 0.23 mm (0.5 mm for water/heptane) and a horizontal gap of 4 mm, which set the detector resolution. The detector slit was 650 mm from the sample.

The counting statistics provided a 0.3% standard deviation for the measurement on the direct beam, 1% at low Qz, 4% at intermediate Qz, 10% at Qz = 0.38, and the largest standard deviation was 15% at Qz = 0.4. As an example of the count rate in the direct beam: 19500 counts were measured in the detector with 110,000 counts in the beam monitor and a background of about 500 counts. A direct beam scan consists of about ten such measurements of the signal and ten measurements of the background. The direct beam is attenuated by a factor of approximately 30,000 compared with most of the other data points. Copper foils are used for this attenuation. They are carefully calibrated to within 1%. Also, the detector linearity is checked and is accurate to within 1% for the counting rates used in this experiment.

## X-ray Analysis

The critical wave vector transfer for total X-ray reflection, $Q_c$ was calculated from the bulk electron densities as $Q_c \approx 4\sqrt{\pi \Delta \rho \, r_e} = 0.01458$ Å$^{-1}$ for the perfluorohexane/water interface and $Q_c = 0.01169$ Å$^{-1}$ for the heptane/water interface, where $\Delta \rho = \rho_{\text{bottom phase}} - \rho_{\text{top phase}}$, $\rho_{\text{water}} = 0.33337$ e–/Å$^3$ at 23.5 ℃, $\rho_{\text{FC6}} = 0.48336$ e–/Å$^3$ at 23.5 ℃, $\rho_{\text{water}} = 0.3333$ e–/Å$^3$ at 25 ℃, $\rho_{\text{heptane}} = 0.2368$ e–/Å$^3$ at 25 ℃, and the classical electron radius $r_e = 2.818$ fm.

**Depletion Layer Model** The model for the depletion layer is derived from the first Born approximation for x-ray scattering, that relates the reflectivity to the electron density gradient normal to the interface averaged over the interfacial plane, $\mathrm{d}\langle \rho(z)\rangle / \mathrm{d}z$, by [13]

$$\frac{R(Q_z)}{R_F(Q_z)} \cong \left| \frac{1}{\rho_{\text{bottom phase}} - \rho_{\text{top phase}}} \int \frac{\mathrm{d}\langle \rho(z)\rangle}{\mathrm{d}z} \exp(-iQ_z z)\mathrm{d}z \right|^2 . \tag{S3}$$

The depletion layer model, uses the electron density profile given by [14]

$$\langle \rho(z)\rangle = \frac{1}{2}\rho_{\text{water}}\left[1 - \mathrm{erf}\left(\frac{z}{\sqrt{2}\sigma_{\text{cap}}}\right)\right] + \frac{1}{2}\rho_{\text{oil}}\left[1 + \mathrm{erf}\left(\frac{z - D_{\text{dep}}}{\sqrt{2}\sigma_{\text{cap}}}\right)\right], \tag{S4}$$

where $D_{\text{dep}}$ determines the distance between midpoints of both water and oil (FC6 or heptane) distributions across the interface and characterizes the drying region at interface. The top and bottom interfaces of depletion layer are assumed to possess equal capillary wave roughness in the main paper, whereas here we present an alternative analysis. The first and second terms in eq. (S3) represent respectively the profile between the water phase and the depletion layer and that between the depletion layer and oil phase. This treatment is analytically equivalent to the monolayer model discussed below, where the electron density of the monolayer $\rho_1$ is fixed to 0 (see also Eq. (S5)).

The x-ray reflectivity calculated from Eq. (S3) using the profile given by Eq. (S4) is shown in Figures 3 and 4 of the main paper. Supplementary Figure 3 demonstrates the results of the depletion model if the interfacial width of the water/vapor and perfluorohexane/vapor internal interfaces in Eq. S4 are assigned different capillary roughness values that are appropriate for the bulk interfaces. Comparison of Supplementary Fig. 3 with Fig. 3 in the main paper shows that the discrepancy is even larger in the Supplementary figure.

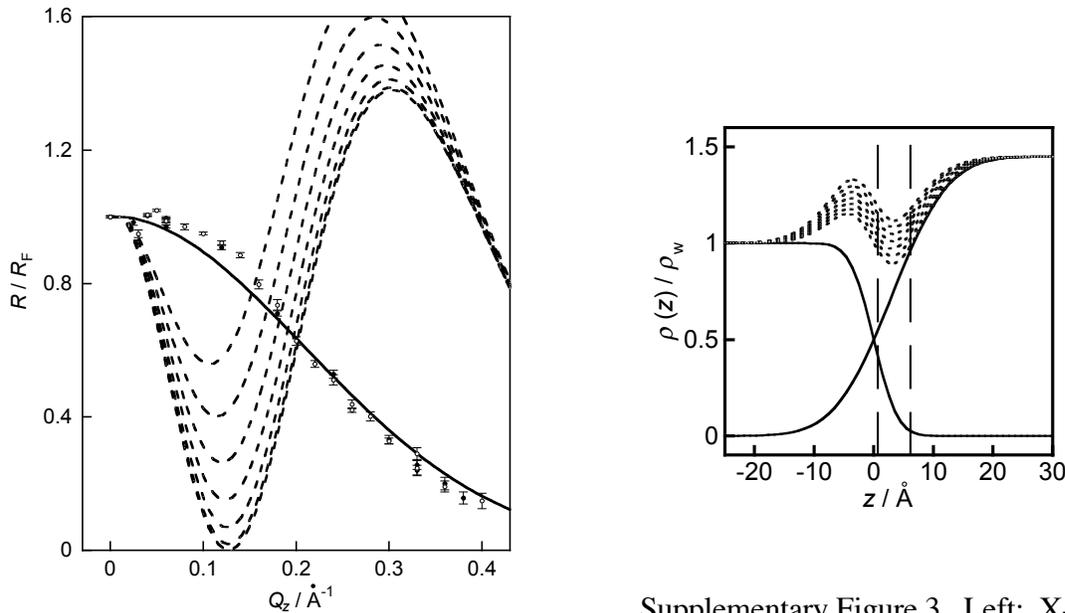

Supplementary Figure 3. Left: X-ray reflectivity normalized to the Fresnel reflectivity analyzed by the depletion layer model for the water/perfluorohexane interface. Symbols are experimental results. Solid line is a fit to Eq.(S1). Dashed lines are determined by Eq. (S4) with depletion distances $D_{\text{dep}} = 0.5$ Å to 3 Å in step of 0.5 Å. This figure is similar to Figure 3 of the main paper except that the capillary wave roughness values for the depletion layer are chosen to correspond to values appropriate for the bulk water/vapor ($\sigma_{\text{cap}} = 3.12$ Å) and bulk perfluorohexane/vapor ($\sigma_{\text{cap}} = 7.48$ Å) interfaces. This figure demonstrates that these bulk values are not the correct choice because the data cannot be fit with them. Right Figure: The electron

density profile normalized to the bulk water, $\rho(z)/\rho_w$, which is used to calculate the corresponding X-ray reflectivity in the figure on the left. The solid lines show the water and perfluorohexane densities separately for the case $D_{dep} = 3$ Å, where there is a 3 Å separation between the midpoints of these lines (indicated by the vertical dashed lines). The curved dashed lines are for $D_{dep} = 0$ Å to 3 Å in steps of 0.5 Å.

**Monolayer Model** We also demonstrate here that the reflectivity data from the perfluorohexane/water interface cannot be adequately modelled by any single layer, not just by a depletion layer. The model applied for monolayer analysis[30] can also be used to examine the deviation around the lower $Q_z$ region of the X-ray reflectivity data by assuming the electron density profile given by,

$$\langle \rho(z) \rangle = \frac{1}{2} \rho_w \left[ 1 - \mathrm{erf}\left( \frac{z}{\sqrt{2}\sigma_{cap}} \right) \right] + \frac{1}{2} \rho_1 \left[ \mathrm{erf}\left( \frac{z}{\sqrt{2}\sigma_{cap}} \right) - \mathrm{erf}\left( \frac{z - L_1}{\sqrt{2}\sigma_{cap}} \right) \right] + \frac{1}{2} \rho_f \left[ 1 + \mathrm{erf}\left( \frac{z - L_1}{\sqrt{2}\sigma_{cap}} \right) \right], \quad (S5)$$

where $L_1$ and $\rho_1$ are the monolayer thickness and electron density as parameters of this model. The top and bottom interfaces of the monolayer have the same capillary wave roughness. The two fitting parameters determined by calculating the X-ray reflectivity from Eq. (S2) are $L_1 = 24^{\pm 3}$ Å and $\rho_1 = 0.4862^{\pm 0.0006}$ e–/Å$^3$, where the fitting results are shown by the long-short dashed lines in Supplementary Figure 4 (Figs. 6a and 6b) together with the prediction from capillary wave theory and multilayer model discussed below. The monolayer thickness is similar to the length of two FC6 molecules perpendicularly oriented to the interface. However, the electron density is much smaller than the value (ca. 0.63 e–/Å$^3$) expected for condensed films (two-dimensional solid phases) of fluorinated compounds such as fluorodecanol (FC10OH) and fluorododecanol (FC12OH) at the C6/water interface [15]. It should be mentioned that the symmetric electron density profile characterized by $L_1 = 24^{\pm 3}$ Å and $\rho_1 = 0.3305^{\pm 0.0006}$ e–/Å$^3$ also provides the same X-ray reflectivity. The integrated electron density deficit can be evaluated as

$D_{int} = \int_0^L (1 - \rho(z)/\rho_w) \mathrm{d}z \approx 8.7 \times 10^{-3} \times 24 = 0.2$ Å [14]. Taking into account that the molecular radius of water is 1.93 Å, it is not likely that the weak electron density deficit extends to a distance of more than five water molecules.  To summarize, this "non-depletion layer" monolayer model does not fit the data properly at high $Q_z$ and does not have a sensible molecular interpretation.

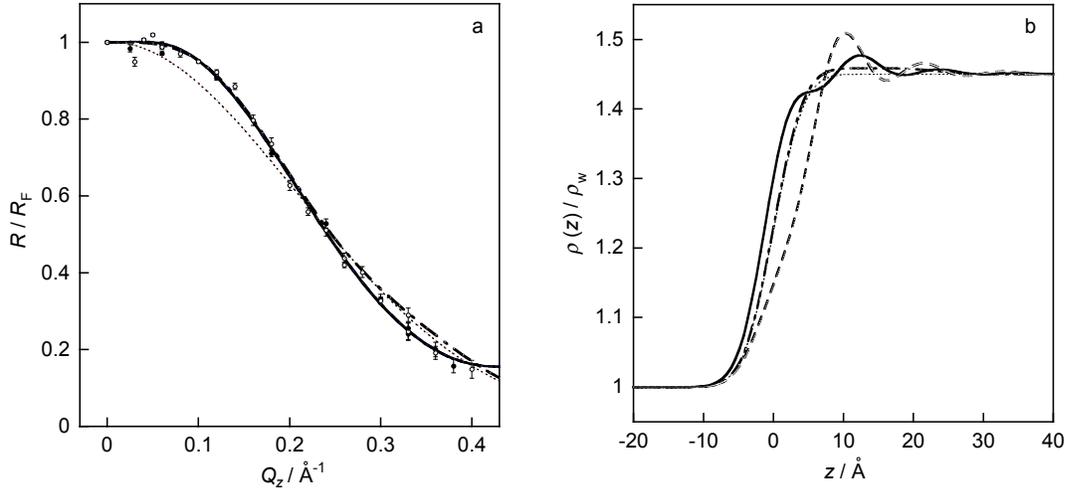

Supplementary Figure 4  Analytical results by multilayer and monolayer models for the water/perfluorohexane interface. (a) X-ray reflectivity normalized to the Fresnel reflectivity and (b) the corresponding electron density profiles normalized to the bulk water: (solid line and short-dashed line) multilayer model; (long-short dashed line) monolayer model; (dotted line) capillary wave theory. The two normalized X-ray reflectivity curves obtained by the multilayer model overlap each other in panel (a).

**Multilayer Model**  The perfluorohexane/water reflectivity data were also analyzed using a multilayer interfacial profile that consists of an exponentially damped cosine. This profile is parameterized by two specific lengths: the oscillation length $l_{osc}$, which is initially fixed to a value corresponding to the length of an FC6 molecule (11.3 Å), and the exponential decay (correlation) length $l_{dec}$ on the FC6 side of the interface. The intrinsic profile $\langle \rho(z) \rangle_{int}$ is expressed as

$$\langle \rho(z) \rangle_{int} = \rho_f + A\left[ C - \cos\left( \frac{2\pi z}{l_{osc}} + \phi \right) \right] \exp\left( -\frac{z}{l_{dec}} \right), \tag{S6}$$

where $A$, $C$, $\phi$, and $l_{dec}$ are fitting parameters. The intrinsic profile is convoluted with a Gaussian function to produce capillary wave roughening. The calculated X-ray reflectivity and the corresponding oscillatory electron density profiles obtained by using Eq. (S6) to fit the measured X-ray reflectivity by Eq. (S3) are drawn as solid and dashed lines in Supplementary Fig. 4 together with the results of the monolayer fit and capillary wave theory. It should be noted that both multilayer profiles yield the same $\chi^2$ values of goodness of fit (and the two fits overlap each other in Figure 4a). These two fits had best fit parameters of (a) $A = 0.76$, $C = 0.09$, $\phi = 243$, $l = 8.32$ A, and (b) $A = 0.38$, $C = 0.14$, $\phi = 169$, $l = 8.32$ A. Subsequent fitting with Eq. (S6) that allowed the value of the oscillation length $l_{osc}$ to vary determined that this parameter is not uniquely specified by these data and that error bars on this length are large. Good fits to the data can be obtained for values of $l_{osc}$ that vary from about half the FC6 molecular length to much longer than the molecular length. The smaller values may be consistent with the FC6 molecules arranged parallel to the interface and the values used in the fitting shown in Supplementary Figure 4 are consistent with FC6 molecules layered normal to the interface. To summarize, this analysis supports the presence of multi-layering of FC6 at the water/FC6 interface, but cannot specify uniquely the details of this layer.

**X-ray Measurement Reproducibility**

Supplementary Figure 5 provides an example of the reproducibility of the reflectivity measurements on the perfluorohexane/water interface.

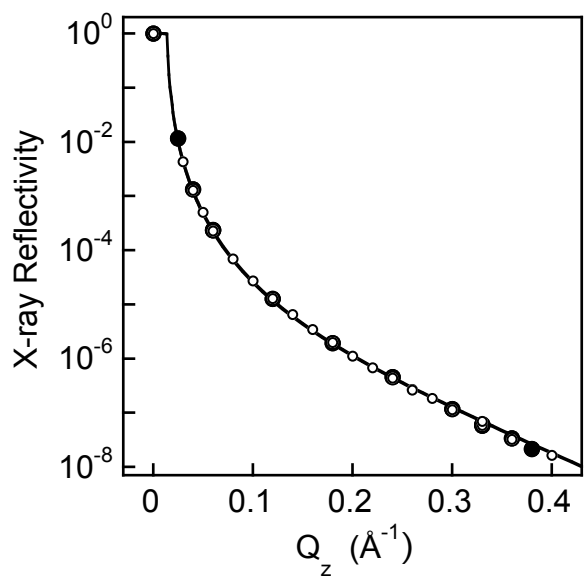

Supplementary Figure 5.  Data shown in Figure 2 of the main paper, but with different symbols for two different samples to indicate the reproducibility of the data.

**Supplementary References**